# Three-dimensional Fermi surface of *2H*-NbSe$_2$ – Implications for the mechanism of charge density waves

*F. Weber[1], R. Hott[1], R. Heid[1], L. L. Lev[2,3], M. Caputo[2], T. Schmitt[2], V. N. Strocov[2]*

[1] *Institute for Solid State Physics, Karlsruhe Institute of Technology, D-76021 Karlsruhe, Germany*
[2] *Paul Scherrer Institut, Swiss Light Source, CH-5232 Villigen PSI, Switzerland*
[3] *National Research Centre "Kurchatov Institute", 123182 Moscow, Russia*

**We investigate the three-dimensional electronic structure of the seminal charge-density-wave (CDW) material *2H*-NbSe$_2$ by soft x-ray angle-resolved photoelectron spectroscopy and density-functional theory. Our results reveal the pronounced 3D character of the electronic structure formed in the quasi-two-dimensional layered crystal structure. In particular, we find a strong dispersion along $k_z$ excluding a nesting-driven CDW formation based on experimental data. The 3D-like band structure of *2H*-NbSe$_2$ has strong implications for the intriguing phase competition of CDW order with superconductivity.**

## 1. Introduction

The interplay between superconductivity and other ground states of solids is one of the most challenging topics in condensed matter physics documented, e.g., in the intense research on high-temperature superconductivity (SC) [1]. It is now evident that a number of layered materials exhibiting charge-density wave (CDW) order, a periodic modulation of the charge carrier density and the atomic lattice, become superconducting once the CDW order is suppressed [2-6]. Classic examples are the members of the transition-metal dichalcogenide family MX$_2$, where M = Nb, Ti, Ta, Mo and X = S, Se, which show a large diversity of CDW ordered phases competing with SC [6,7].

*2H*-NbSe$_2$ is special among the transition-metal dichalcogenides in that it supports CDW order ($T_{CDW}$ = 33 K [8,9]) and superconductivity ($T_c$ = 7.2 K [10]) simultaneously at ambient pressure and without intercalation or chemical substitution. For decades, the driving mechanism of the CDW transition was hotly debated including various ARPES studies [11-16] searching for the Fermi surface (FS) nesting required for a 2D analogue of the Peierls transition [17,18]. While some reports on ARPES experiments supported the FS nesting, i.e., weak coupling scenario, others did not. On the other hand, the strong coupling model with wave vector dependent electron-phonon coupling (EPC) matrix elements, originally introduced in the late 1970s [19,20], could recently explain many of the puzzling results from a broad range of experiments [16,21-24].

In our work, we used soft x-ray angle-resolved photoemission spectroscopy (SX-ARPES) in order to unveil the three dimensional (3D) band structure and the corresponding FS of *2H*-NbSe$_2$. The increase of the mean free path of photoelectrons with their kinetic energy results in their stronger delocalization in the out-of-plane direction and therefore, by the Heisenberg uncertainty principle, in higher intrinsic definition of the out-of-plane momentum $k_z$ [25,26]. We compare our experimental data to *ab-initio* band structure calculations based on density-functional theory (DFT) and find good overall agreement. Our results demonstrate that *2H*-NbSe$_2$ features no FS nesting related to the CDW formation. Hence, this report finally solves the long-standing debate about the electronic structure of *2H*-NbSe$_2$ in favor of the strong electron-phonon coupling scenario. Further, our result provides an indispensable starting point to understand the phase competition between the CDW order and SC not only in bulk, but also thin sheets and single layers of NbSe$_2$, where the latter have been studied recently as model systems for the effect of dimensionality on competing electronic phases [27,28].

## 2. Experimental and computational details

SX-ARPES experiments were performed on in-situ cleaved single crystals of *2H*-NbSe$_2$ at the SX-ARPES endstation (for details see [29]) at ADRESS beamline [30] of the Swiss Light Source (SLS) with incident photon energies *hν* in the range 600-850 eV. The combined (beamline and analyzer) experimental energy resolution varied, correspondingly, from 85 to 120 meV for FS measurements, and for the electron band structure images it was about 50 meV. The measurements were performed at low sample temperature $T$ = 11 K in order to suppress relaxation of **k**-selectivity because of thermal motion [31]. Density functional theory (DFT) calculations were carried out in the framework of the mixed basis pseudopotential method [32] using the local density approximation (LDA) (for details see [22]). All results presented here were obtained for the experimental hexagonal structure ($a = b =$ 3.44 Å, $c$ = 12.5 Å) [33].

## 3. Results: Experimental and theoretical electronic structure

Before we discuss the FS geometry in detail, we focus on electronic band dispersions in comparison with DFT calculations along different directions of the Brillouin zone (BZ) depicted in Figure 1(a). Figures 1(b,c) show experimental ARPES images along the ΓM direction of the BZ. The measurements were performed at photon energy *hν* = 632 eV corresponding to the out-of-plane electron momentum $k_z$ =0. Here we have used *p*- and *s*-polarized x-rays in order to separate electronic states having different symmetries relative to the ΓALM symmetry plane. The ARPES image obtained with *p*- polarization [Fig. 1(b)] reveals the symmetric electron states which form a band manifold extending from the Fermi energy $E_F$ down to a binding energy $E_B$ of -4 eV. These bands include, in particular, those crossing $E_F$ to form the FS. The corresponding DFT calculation [solid blue lines in Fig. 1(b,c)] predicts three hole-like FS pockets around the $\Gamma$ point, with the innermost "pancake" one formed by the 3D Se $4p_z$ orbitals, and the two outer ones by the quasi-2D bonding and antibonding Nb $d_{3z^2-r^2}$ orbitals. In turn, the ARPES image (c) obtained with *s*-polarized x-rays reveals the electron states antisymmetric relative to the $\Gamma ALM$ plane which are centered around $E_B = -2$ eV. In the following, we will only discuss results obtained with *p*-polarization because we are mostly interested in the bands forming the FS. The ARPES image measured at $h\nu = 632$ eV with *p*-polarization to reveal electron states symmetric relative to the $\Gamma AHK$ plane is shown in Figure 1(d) also superimposed with corresponding DFT calculations. Generally, we find a good agreement with the DFT calculation over the whole energy range. We note that we did not apply any shift of the Fermi level or other scaling which is often done in such comparisons of *ab-initio* calculated and measured band structures. Regarding the known problems of DFT for materials with strong electronic correlations this confirms that *2H*-NbSe$_2$ is only weakly correlated.

An important aspect of the electronic structure of *2H*-NbSe$_2$ is whether the 3D Se $4p_z$ bands cross $E_F$ to form the 'pancake' FS pocket around the $\Gamma$ point. If the FS nesting scenario of the CDW formation would apply (which is however ruled out by the present work, see below) the spectral weight of Nb $d_{3z^2-r^2}$ states at $E_F$ would have been removed by the CDW gap and the Se $4p_z$ states would have been the only remaining electronic states at the Fermi level for superconductivity in *2H*-NbSe$_2$ [e.g. see Ref. [34]]. The experimental data in Figure 1(b,d) seem to indicate a dispersion of the Se $4p_z$ derived band with a maximum appearing slightly below $E_F$. However, this downward shift of the ARPES intensity can in fact result from $k_z$ broadening of the ARPES final states combined with finite experimental resolution which may shift the ARPES peaks near band extremes toward the band interior. Analyzed in detail in [25,35], this effect was recently observed and reproduced by one-step photoemission calculations, for example, for Ba$_{1-x}$K$_x$Fe$_2$As$_2$ [36]. The question of whether the Se $4p_z$ band crosses $E_F$ is resolved by the ARPES data shown in Figure 1(e,f) taken for the $\Gamma M$ direction of the BZ at $h\nu = 732$ eV. Formally, this image corresponds to the same $\Gamma M$ direction ($k_z = 0$) as in Figure 1(*b*), but we see here a different set of bands (in the following we discuss that the difference in ARPES response is related to fundamental symmetry group properties of *2H*-NbSe$_2$). Most important, a zoom-in of the ARPES intensity in vicinity of the $\Gamma$ point [Fig. 1(f)] clearly demonstrates that the band formed by the 3D Se $4p_z$ orbitals crosses $E_F$ to form the 3D "pancake" FS pocket, closing this important issue in the electronic structure of *2H*-NbSe$_2$. This fact is again in agreement with DFT predicting the band crossing of $E_F$.

Also, we note that whereas our DFT calculations correctly reproduce energy position of the Nb $d_{3z^2-r^2}$, the calculated Se $4p_z$ states appear somewhat high in energy. The difference in DFT predictions between these two different types of electron states may come from difference in self-energy corrections connected with different character of the corresponding wavefunctions [37].

We will now turn to discussion of the FS as observed with *p*-polarized x-rays. Figures 2(a,b) show ARPES intensities at $E_F$ in the $k_x$-$k_y$ $\Gamma KM$ plane of the BZ corresponding to formally equivalent $k_z$ values in the $\Gamma_{26}$ ($h\nu = 634$ eV) and $\Gamma_{27}$ (674 eV) points. As we discuss below, the difference between these images reflects the non-symmorphic space group of *2H*-NbSe$_2$. All three hole pockets around the $\Gamma$ point are well resolved in the cut including $\Gamma_{26}$ (corresponding to $k_z \approx 13$ Å$^{-1}$) along with the outer FS cylinder around the *K* point [Fig. 2(a)]. Overall, the shape of the two outer Nb $d_{3z^2-r^2}$ pockets around $\Gamma$ is well reproduced by DFT, which however, slightly underestimates their size. In contrast, the size of the Se $4p_z$ inner pocket is largely overestimated. However, we remind the reader that, experimentally, this inner-most pocket intensity falls slightly below $E_F$ [see Fig. 1(d)] due to the combined $k_z$-broadening and energy resolution effects. Whereas the outer *K* pocket is clearly visible in the $\Gamma_{26}$ cut [Fig. 2(a)], the inner one can be observed in the $\Gamma_{27}$ cut with $k_z \approx 13.5$ Å$^{-1}$ [Fig. 2(b)]. Band degeneracy at the zone boundary along the $c^*$ direction, i.e. $k_z$ at the A point achieved at $k_z \approx 12.75$ Å$^{-1}$ ($h\nu = 609$ eV), results in the observation of only two FS pockets, one around *A* and one around *K* [Fig. 2(c)].

Results shown in Figure 3 focus on the $k_z$ dependence of the FS in three different $k_y$ cuts with the value of $k_y$ indicated by the dashed lines in Figure 3(a). While two lines include the high symmetry points $\Gamma$ and *K* where the FS pockets are centered (see Fig. 2), the third line was chosen based on a previous ARPES study [14] claiming potential FS nesting (see below).

In FS cut along the $M\Gamma M$ line [Fig. 3(b)] the $k_z$ values corresponding to the $\Gamma$ points can be best identified by the intensity of the 3D Se $4p_z$ band at $E_F$ with its clear $k_z$ dependence. The $k_z$ dispersion of the outer $\Gamma$ pockets is less pronounced but visible, e.g., for 12.5 Å$^{-1} \leq k_z \leq 13$ Å$^{-1}$. Interestingly, comparison to calculations shows that the Se $4p_z$ intensity can only be observed at every fourth zone center. For the most common materials, whose space group is symmorphic, the ARPES intensity should replicate every BZ, and for *2H*-NbSe$_2$ belonging to more rare materials having non-symmorphic space group such as graphite [38], WSe$_2$ [39], WTe$_2$ [40] and CrO$_2$ [41] the intensity should replicate every second BZ. The fourfold reduced $k_z$-periodicity observed in our case evidences a strong impact of ARPES matrix elements beyond the basic symmetry considerations. This effect is further highlighted by the experimental $k_z$-dispersion map along the $\Gamma$-A line, i.e., at $k_x = 0$ [Fig. 3(c)]. ARPES intensity from the antibonding and bonding Se $4p_z$ bands form a systematic dispersion pattern, where the former reaches $E_F$ at $k_z \approx 13$ Å$^{-1}$ ($\Gamma_{26}$ point) and again at $\approx 15$ Å$^{-1}$ ($\Gamma_{30}$), and the latter reaches its dispersion minimum at $k_z \approx 14$ Å$^{-1}$ ($\Gamma_{28}$). This $k_z$-dispersion pattern explains why the ARPES images in Fig. 1 (b) and (e,f) show different set of bands. At $k_z \approx 13.5$ Å$^{-1}$ ($\Gamma_{27}$ point) and $\approx 14.5$ Å$^{-1}$ ($\Gamma_{29}$) the ARPES intensity sharply switches between the antibonding and bonding Se $4p_z$ bands that visually forms one continuous band whose periodicity is fourfold reduced compared to the formal unit cell. To the best of our knowledge, we here report the first observation of this unexpected ARPES matrix element effect which calls for a solid theoretical analysis based on the fundamental crystal symmetry properties. Interestingly, the FS map in Fig. 3(b) shows that this unexpected symmetry effect is restricted to the Γ-A direction, returning to the twofold periodicity reduction characteristic of the non-symmorphic symmetry group. We note that in the ARPES image measured at $hv = 732$ eV [Fig. 1(e,f)] the Se $4p_z$ intensity is suppressed near the Γ point because of the fourfold periodicity reduction, but away from this point it emerges to allow our clear identification of the $E_F$ crossing described above [see Figs. 1(e,f)].

In FS cut along the $K$-$K$ line [Fig. 3(d)] the FS pockets centered on the $K$ point display again a pronounced $k_z$ dependence. DFT predicts degenerate bands in the plane containing the A point, i.e., the zone boundary along the $c^*$ direction, and shows the largest difference in the $\Gamma$ plane. This is confirmed by the observed intensities again with a notable impact of matrix elements expected from the non-symmorphic space group of 2H-NbSe$_2$: the dominating intensity changes alternatingly between the bands corresponding to the inner and outer $K$ pockets for successive $\Gamma$ planes [e.g. see Fig. 3(d) in the range 0.5 Å$^{-1}$ ≤ $k_x$ ≤ 0.75 Å$^{-1}$]. Note that the broad intensity enhancement in the $k_x \approx 0$ region originates from electronic bands located outside the $K$-$K$ line whose signal is detected because of the finite experimental angular resolution of the experiment in the $k_y$ direction (perpendicular to the analyzer slit) and imperfection of the sample surface planarity. This stray intensity can be seen easily in Figure 2. The DFT calculations, naturally having perfect **k**-resolution, predicts no spectral intensity strictly along this line. The same effect is actually responsible for the signal around $k_x \approx \pm 1$ Å$^{-1}$ in Fig. 3(b) although its signature is weaker since that dataset was taken with a higher momentum resolution, i.e. smaller analyzer slit.

Overall, we conclude that our DFT calculations well reproduce the full 3D FS of 2H-NbSe$_2$ indicating a weakly correlated nature of this material.

## 4. Results: CDW mechanism

Static CDW order, *i.e.*, a periodic modulation of the electronic density, reflects an enhancement of the dielectric response of the conduction electrons at the CDW wavevector, $q_{CDW}$, but it has long been known that it is only stabilized by a coupling to the crystal lattice [18,42]. Transitions into the CDW phase on lowering the temperature are accompanied by a softening of an acoustic phonon at $q_{CDW}$ to zero frequency at $T_{CDW}$ where it freezes into a static distortion [43] and evolves into the new periodic (often incommensurate) superstructure. Chan and Heine derived the criterion for a stable CDW phase with a modulation wavevector $q$ as [42]

$$\frac{4\eta_q^2}{\hbar\omega_q} \geq \frac{1}{\chi_q} + (2U_q - V_q)$$

where $\eta_q$ is the **k**-integrated electron-phonon coupling matrix element associated with a mode at an energy of $\omega_q$, $\chi_q$ is the dielectric response of the conduction electrons, and $U_q$ and $V_q$ are their Coulomb and exchange interactions. Although both sides of this inequality are essential in stabilizing the CDW order, the common assumption is that the ordering wave vector, $q_{CDW}$, is determined by the right-hand side, *i.e.*, by a singularity in the electronic dielectric function $\chi_q$ at $q_{CDW}$.

Our SX-ARPES study derives the 3D FS based on detailed comparison to experimental data and, thus, solves any remaining doubts about the electronic structure. We will now return to analysis of our results in the context of CDWs. We focus on the FS cut along the lowest line across the BZ, indicated in Fig. 3(a). This selection is based on previous ARPES publications [14,15], where it was reported that the CDW ordering wave vector $q_{CDW}$ = (0.329,0,0) connects parts of the inner K pocket along this line. Assuming a quasi-2D behavior, the authors of Ref. [14] interpreted their results in favor of a FS nesting driven CDW formation in 2H-NbSe$_2$. However, our results in Fig. 3(e), both theory and experiment, demonstrate that there is a substantial band dispersion along $k_z$ for this in-plane direction. Again, the two bands are degenerate at $k_z$ values corresponding to zone boundaries along $k_z$, i.e., $k_z \approx 12.75$ Å$^{-1}$, 13.25 Å$^{-1}$, 13.75 Å$^{-1}$. Going to intermediate $k_z$ values, corresponding to those of $\Gamma$ points, we see a pronounced dispersion of the inner $K$ pocket. This observation rules out any significant nesting of the FS. The FS dispersion corresponding to the outer $K$ pocket also show pronounced $k_z$ dependence. Experimental data are in good agreement with DFT exhibiting again intensities distributed alternatingly on the two bands at $k_z \approx 13$ Å$^{-1}$, 13.5 Å$^{-1}$, 14 Å$^{-1}$. No nesting can be seen in the calculated band structure.

Our results on the FS nesting are seemingly at odds with previous low-energy ARPES measurements with $hv = 23$ and 55 eV [15] and 50 eV [14] which reported well-defined hotspots on the inner $K$ pocket connected by $q_{CDW}$ [15] that were interpreted in [14] as signatures of the FS nesting being involved in the formation of the CDW. However, we note that the inner $K$ pocket shifted by $q_{CDW}$ intersects at an angle of 120° with the original $K$ pocket. Therefore, the CDW ordering wave vector connects only isolated points of the Fermi surface and no nesting is present in the ($k_x$,$k_y$) plane. Furthermore, our results [see Figs. 3(e) and 4] show that these hotspots are restricted to a narrow range of $k_z$ values near the $\Gamma$ points, and the FS contours for generic $k_z$ values are not connected by the same wave vector anymore. Hence, the hotspots in the ($k_x$,$k_y$) plane do not correspond to lines along $k_z$ connected by $q_{CDW}$ as it would be expected for a quasi-2D material with a corresponding 2D, i.e. $k_z$-independent, FS topology. The conclusions of the previous ARPES works about the FS nesting were misled by a particular choice of the $hv$ values bringing $k_z$ close to the $\Gamma$ point. Our analysis of the SX-ARPES data supported by DFT calculations allows us to conclude that 2H-NbSe$_2$ is a seminal example where the FS nesting is not only irrelevant for the CDW formation but altogether absent.

The experimental evidence about the absence of FS nesting in 2H-NbSe$_2$ are supported by analysis of our DFT results. To visualize the full momentum dependence, we plot the complete calculated FS in Figure 4(a). White arrows indicate the FS parts on the inner K pocket, which are connected by $q_{CDW}$. No nesting is present and this is particularly clear for the inner K pocket [dark green FS body in Fig. 4(a)]. Finally, we deduce the electronic joint density-of-states (eJDOS) closely related to the imaginary part of the electronic susceptibility and, thus, the nesting function [Fig. 4(b)]. This quantity integrates over all electronic states, and its flat momentum dependence demonstrates that there is no tendency towards FS nesting at $q_{CDW}$ at all.

The alternative strong EPC model of the CDWs introduced in the late 1970s [19,20] got recently more support [16,21-24] for 2H-NbSe$_2$. Previous works [21,22] have shown that lattice dynamical calculations based on analogous band structure calculations reproduce well the observed phonon softening, and the momentum dependence of $\eta_q^2$ could be determined within these calculations. The analysis of our SX-ARPES experiment demonstrates that in addition to the lattice dynamical properties the DFT is able to quantitatively describe the electronic band structure of 2H-NbSe$_2$. Crucial for the relevance of the DFT is the small strength of many-body interactions in this material.

With the FS nesting mechanism ruled out as irrelevant for 2H-NbSe$_2$, the CDW ordering wave vector is therefore determined by the momentum dependence of the EPC matrix elements. Apart from recent *ab-initio* calculations, Doran already considered the impact of wave vector dependent EPC matrix elements on determining the CDW ordering wave vector in 2H-NbSe$_2$ in the late 1970s [19]. Based on the work of Varma and Weber [44,45], he calculated the contribution $D_2$ to the dynamical matrix, which effectively describes the electronic polarization effects, i.e., the combination of EPC matrix elements $\eta_{k,k+q}^2$ and the bare electronic response function,

$$D_2 \propto \sum_k \eta_{k,k+q}^2 \frac{f_k - f_{k+q}}{E_{k+q} - E_k},$$

where the $f$'s are Fermi functions. Assuming that the potential is carried rigidly by the displaced ions and that there is only a single symmetric band, the EPC matrix element $\eta_{k,k+q}^2$ can be expressed in terms of the Fermi velocities $v_k^\alpha = \partial E / \partial k^\alpha$ ($\alpha$ being the direction of $q$) [19]

$$\eta_{k,k+q}^2 \propto \left| v_k^\alpha - v_{k+q}^\alpha \right|^2.$$

Thus, he could demonstrate that $D_2$ exhibits a peak around $q \approx (0.3,0,0)$, which is in reasonable agreement with the observed $q_{CDW}$ regarding the severe approximations, whereas the bare electronic susceptibility is featureless along the $\Gamma$-M direction in 2H-NbSe$_2$ [46]. We can illustrate this result based on our calculated FS bodies. To this end, we consider only the FS body corresponding to the inner K pocket because ARPES identified CDW hotspots only on this FS part [14,15]. The momentum dependence of the electronic joint density of states can be interpreted as the overlap integral of the two FS bodies shifted by a wave vector $q \parallel q_{CDW}$ [within the vertical plane indicated in Fig. 5]. It is clear that the overlap is complete/maximum for $q = 0$ and is going monotonically down to zero at $|q| = |q_{max}|$, where $q_{max}$ is the maximum diameter of the FS body parallel to $q$. Hence, $D_2$ will be zero for $|q| \geq |q_{max}|$ as well. In the second step we consider the difference of the Fermi velocities $\Delta v$ at the intersection points of the shifted FS bodies. The illustration in Figure 5 shows that the difference between $v_k$ and $v_{k+q}$ indicated by the horizontal arrow is parallel to $q$ and maximum for $|q| = |q_{max}|$. Hence, the product of the decreasing electronic jDOS and the increasing difference of the Fermi velocities (as function of $q$) will determine the position of the maximum of $D_2$, and hence of $q_{CDW}$, in momentum space. Here, we emphasize the conclusion of Doran [19] that the CDW ordering wave vector is not simply related to the Fermi surface geometry and one should not expect to see anything from drawing so-called spanning vectors in it.

The irrelevance of the FS nesting mechanism has direct consequences not only for the competition between CDW ordered and superconducting phases in 2H-NbSe$_2$ but also for our understanding of the evolution of the respective transition temperatures $T_{CDW}$ and $T_c$ as function of dimensionality. First, the fact that the CDW gaps only a tiny part of the FS and leaves the remaining large FS parts for superconductivity explains the easy coexistence of CDW and superconductivity: suppressing CDW order frees only a very small part of the FS for superconductivity. Also the suppression of $T_{CDW}$ by pressure or intercalation has only a small impact on $T_c$ [16,47,48]. On the other hand, $T_{CDW}$ ($T_c$) is enhanced (reduced) by roughly a factor of three (two) in single sheets of NbSe$_2$ [28]. We can understand this behaviour because single sheets are for all practical aspects 2D materials featuring a 2D FS. Consequently, the hotspots will become hotlines and CDW formation in single sheets of NbSe$_2$ is likely supported by both, EPC matrix elements and FS nesting.

## 5. Conclusion

In summary, we have reported a combined SX-ARPES and DFT study of the electronic band structure of 2H-NbSe$_2$. A complex space group of this material results in its complex ARPES response. For the first time we accurately derive the full 3D topology of the FS of 2H-NbSe$_2$ and, thereby, resolve the decade long debate on the origin of CDW formation in this seminal material with the results that FS nesting is irrelevant. The precise knowledge of the FS provides a direct understanding of the phase competition between CDW order and superconductivity in bulk *2H*-NbSe$_2$ but also for the effects of dimensionality on correlated electronic phases.

**Acknowledgements:**
F.W. was supported by the Helmholtz Society under contract VH-NG-840. M.C. acknowledge funding from the Swiss National Science Foundation under the grant No. 200021_165529.

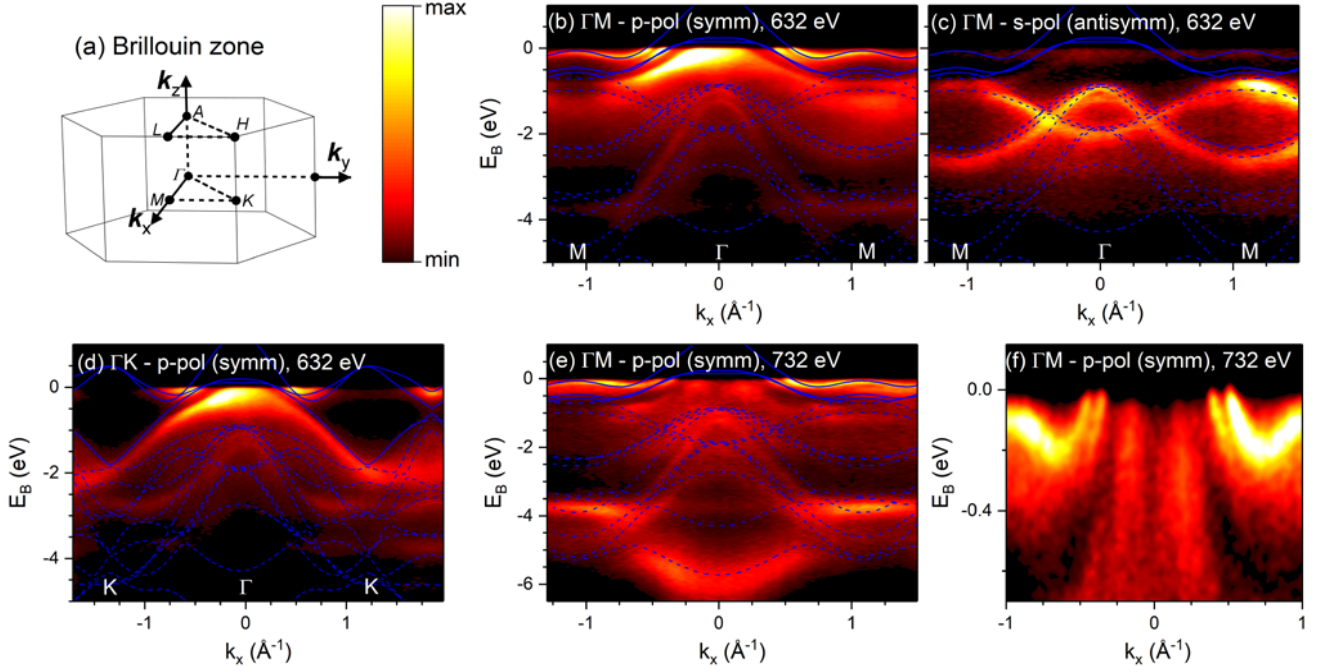

**FIG. 1.** (a) BZ of *2H*-NbSe$_2$. Capital letters denote high symmetry points. (b)(c) Comparison of calculated band structure (dashed lines) and SX-ARPES intensities measured at $h\nu = 632$ eV along the $\Gamma - M$ direction distinguished for (b) symmetric and (c) antisymmetric states employing *p*- and *s*-polarized light (see text). Two Nb $d_{3z^2-r^2}$ and one Se $p_z$ bands making up the Fermi surface are shown as solid lines. (d) The same comparison along the $\Gamma - K$ direction for the symmetric states employing *p*-polarized light (see text). (e) Same direction and *p*-polarization as in (b) but measured with higher $h\nu = 732$ eV, corresponding to $\Gamma_{28}$, and (f) its zoom-in around the $\Gamma$ point. The Se $p_z$ bands clearly cross $E_F$ to form the 3D "pancake" FS pocket.

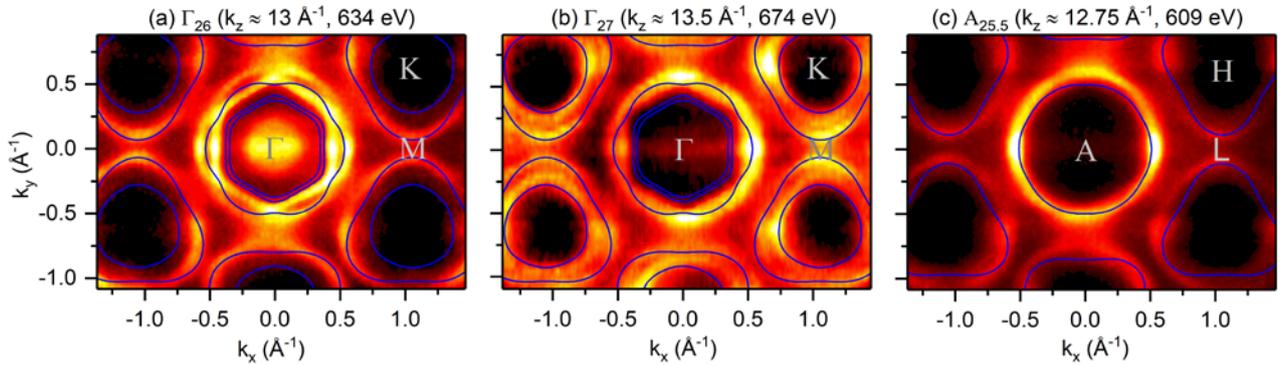

**FIG. 2.** Comparison of calculated Fermi surface (lines) and SX-ARPES intensities observed in the $\Gamma$KM plane of the BZ with (a) $k_z \approx$ 13 Å$^{-1}$ achieved at $h\nu = 634$ eV and (b) 13.5 Å$^{-1}$ at 674 eV, and (c) intensities in the ALH plane of the BZ with $k_z \approx$ 12.75 Å$^{-1}$ at $h\nu =$ 609 eV. High-symmetry points are indicated by (grey) capital letters.

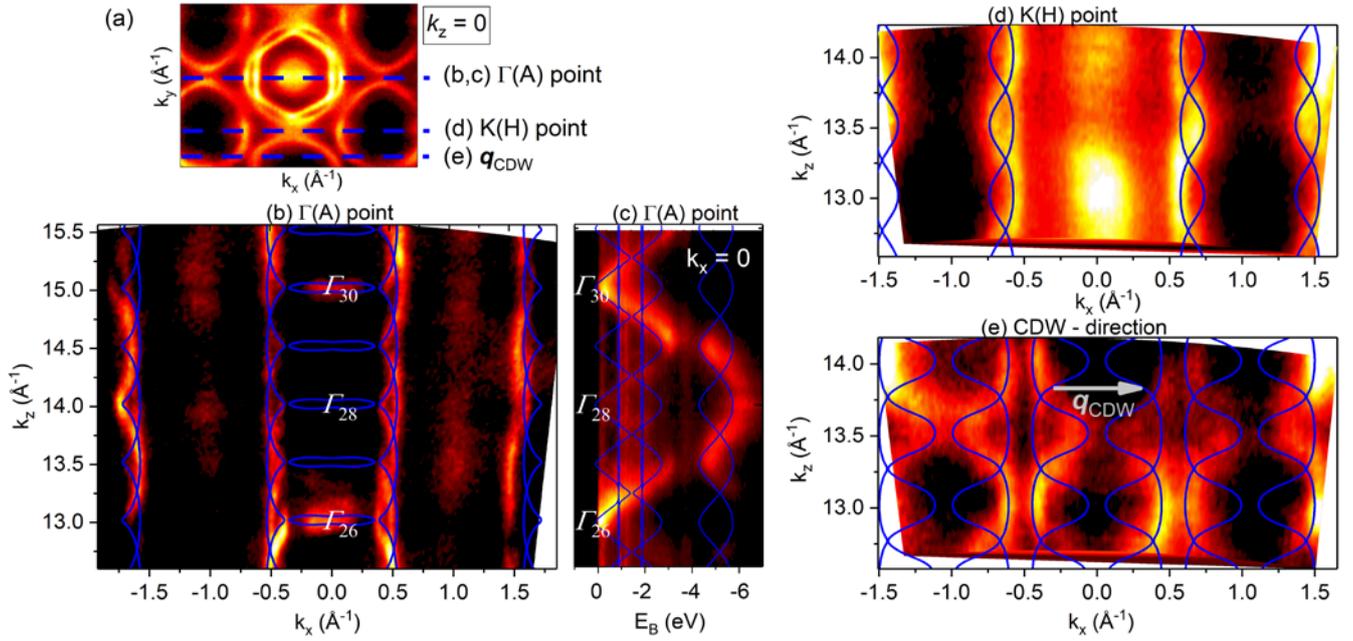

**FIG. 3.** Fermi surface in $k_x - k_z$ planes for three different values of $k_y$ indicated by the dashed lines in (a) plot of intensities in the $k_x - k_y$ plane ($k_z \approx 13$ Å$^{-1}$) [same data as in Fig. 2(a)]. (b)(d)(e) Comparison between calculated band structure (lines) and observed SX-ARPES intensities for $k_x - k_z$ planes including (b) the Γ(A) point, (d) the K(H) point, and (e) for the direction indicated in (a) which was proposed to be relevant for FS nesting in Ref. [14]. In (e), the ordering wave vector $q_{CDW}$ (grey arrow) connects the FS dispersions only in a narrow $k_z$ range and can therefore not be related with any FS nesting. (c) The band dispersion along the Γ - A line [corresponding to $k_x = k_y = 0$ in panel (b)] that shows the fourfold reduced $k_z$-periodicity.

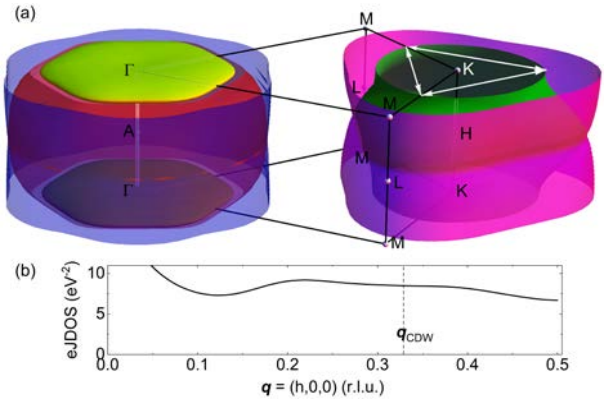

**FIG. 4.** (a) Calculated 3D FS. Capital letters denote high symmetry points of the Brillouin zone. White arrows connect parts of the inner K pocket, where nesting was proposed [14] and CDW hotspots were reported [15]. (b) Calculated electronic joint density-of-states (eJDOS) corresponding to the FS shown in (a).

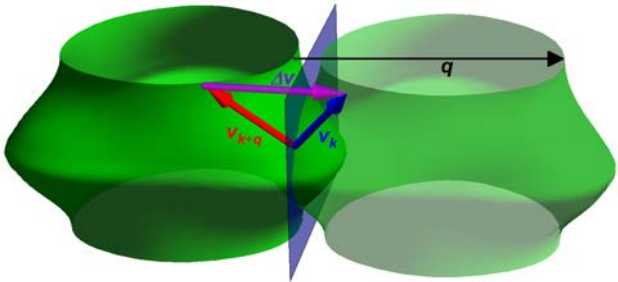

**FIG. 5.** Illustration of the model of Doran [19] considering the FS body corresponding to the inner K pocket. $q$ denotes the wave vector by which the FS body on the right is shifted with regard to the one on the left. $v_k$ and $v_{k+q}$ and the corresponding arrows indicate the Fermi velocities at one intersection point of the original and shifted FS bodies, respectively, and $\Delta v = v_k - v_{k+q}$.